\documentclass[aps,prd,twocolumn,superscriptaddress,preprintnumbers,showpacs,nofootinbib]{revtex4}

\usepackage{graphicx}
\usepackage{hyperref}

\newcommand{\bbar}{\bar{\beta}}
\newcommand{\abs}[1]{\left| #1\right|}
\newcommand{\thetae}{\theta_{\mathrm{E}}}

\newcommand{\cmwe}{\hbox{ cmwe}}

\newcommand{\murad}{\hbox{ $\mu$rad}}
\newcommand{\ev}{\hbox{ eV}}

\newcommand{\mev}{\hbox{ MeV}}
\newcommand{\gev}{\hbox{ GeV}}

\newcommand{\m}{\hbox{ m}}
\newcommand{\km}{\hbox{ km}}
\newcommand{\pc}{\hbox{ pc}}
\newcommand{\kpc}{\hbox{ kpc}}
\newcommand{\mpc}{\hbox{ Mpc}}
\newcommand{\kg}{\hbox{ kg}}
\newcommand{\s}{\hbox{ s}}
\newcommand{\y}{\hbox{ y}}
\newcommand{\my}{\hbox{ My}}
\newcommand{\cfrac}[2]{\textstyle{\frac{#1}{#2}}}
\newcommand{\nucl}[2]{\ensuremath{\;^{#2}\hbox{#1}}}
\newcommand{\tnucl}[2]{\ensuremath{^{#2}\hbox{#1}}}

\begin{document}

\preprint{FERMILAB--PUB--06/051--T}
\preprint{Roma-TH-1437}

\title{Gravitational Lensing of Supernova Neutrinos}


\author{Olga Mena}
\email[]{Olga.MenaRequejo@roma1.infn.it}
\affiliation{Theoretical Physics Department, Fermi National 
Accelerator Laboratory, P.O.~Box 500, Batavia, Illinois 60510 USA}
\affiliation{ INFN Sez.\ di Roma,                     
Dipartimento di Fisica,                 
Universit\`{a} di Roma``La Sapienza'',       
P.le A.~Moro, 5,                        
I-00185 Roma,
Italy}
\author{Irina Mocioiu}
\email[]{irina@phys.psu.edu}
\affiliation{Department of Physics, Pennsylvania State University, University 
Park, Pennsylvania 16802-6300 USA}
\author{Chris Quigg}
\email[]{quigg@fnal.gov}
\affiliation{Theoretical Physics Department, Fermi National 
Accelerator Laboratory, P.O.~Box 500, Batavia, Illinois 60510 USA}


\date{\today}
\begin{abstract}
The black hole at the center of the galaxy is a powerful lens for 
supernova neutrinos. In the very special circumstance of a supernova 
near the extended line of sight from Earth to the galactic center, 
lensing could dramatically enhance the neutrino flux at Earth and 
stretch the neutrino pulse.
\end{abstract}


\maketitle

\section{Introduction}
Surely neutrinos---in common with other forms of matter and
energy---experience gravitational interactions~\cite{Brill}.  Where is
the observational evidence to support this assertion?  No analogue of
the classic Einstein~\cite{Albert}--Eddington~\cite{Eclipse}
demonstration of the deflection of starlight by the Sun is in prospect.
No continuous intense point source of neutrinos is known, and the
angular resolution of neutrino telescopes---a few degrees achieved at
Super-Kamiokande in the few-MeV range and approximately
$\cfrac{1}{2}^{\circ}$ projected for km$^{3}$-scale ultrahigh-energy
neutrino telescopes---is poorly matched to the anticipated
1.75-arcsecond deflection of neutrinos from a distant source.
Accordingly, we must look elsewhere.

Neutrino oscillations arise from phase differences in the propagation
of different \emph{inertial-}mass eigenstates.
Equivalence-principle--violating models---massless or mass-degenerate
neutrinos, with gravity coupling nonuniversally to different
flavors---give a poor description of neutrino-oscillation
phenomena~\cite{Mann:1995nw}.

The arrival time of the Supernova 1987A neutrino burst, recorded within
three hours of the associated optical display after a $166\,000$-year voyage,
argues that neutrinos and photons follow the same trajectories in the
gravitational field of our galaxy, to an accuracy better than
0.5\%~\cite{Longo:1987gc,Krauss:1987me}.  For a selection of modern 
halo profiles for our galaxy~\cite{Ascasibar:2005rw}, the Shapiro time 
delay~\cite{PhysRevLett.13.789} for photons arriving from SN1987A 
ranges between $(9.26\hbox{ -- }11.5) \times 10^{6}\s = 0.29\hbox{ -- }0.36\y$. 
[Observe
that, for small neutrino masses, neutrinos lag light by only $\Delta t
\approx \cfrac{1}{2}(m_{\nu}/p_{\nu})^{2}t \approx 2.5~\mu\mathrm{s}
\cdot (m_{\nu}/1\ev)^{2}/(p_{\nu}/1\gev)^{2}$ over the SN1987A--Earth
trajectory.]

Even if they experience normal gravitational interactions, neutrinos do
not cluster gravitationally on small scales, because of their large
velocities.  {Free-streaming} neutrinos inhibit the growth of
density fluctuations, and so leave an imprint on the
large-scale-structure matter-power spectrum that is directly related to
the neutrino energy density~\cite{Cooray:1999rv}.  Weak-lensing surveys
offer the most promising path to precise measurements of the
matter-power spectrum and might, in the future, be sensitive to a
nonzero {(inertial)} neutrino
mass~\cite{Kaplinghat:2003bh,Lesgourgues:2005yv,Lesgourgues:2006nd}. 

The SN1987A argument, though telling, is indirect. Can we imagine more 
direct manifestations of gravity's influence on neutrinos? 

If it could be carried out, a neutrino analogue~\cite{Raghavan:2006xf}
of the Pound--Rebka experiment~\cite{PoundRebka}, applying the
M\"{o}ssbauer effect to recoilless resonant capture of
antineutrinos~\cite{PhysRev.116.1581,PhysRevC.28.2162}, would
demonstrate the blue shift of neutrinos falling in a gravitational
field.

In this Article, we explore the possibility that gravitational lensing 
of neutrinos could be observed in special circumstances. Neutrinos 
emitted by a supernova on the other side of our galaxy would be lensed 
by the black hole at the galactic center. In the exceedingly rare circumstance 
of source--lens--observer syzygy, the flux of neutrinos 
arriving at Earth would be amplified by many orders of magnitude. For 
the somewhat less improbable case of near-perfect opposition, lensed 
neutrinos would reach Earth by paths of differing lengths, resulting 
in an observable time dispersion of the supernova neutrino burst. 

After computing the amplification and time dispersion for various
configurations, we estimate the rate at which appropriately
providential circumstances might occur.  We assess the diffusing
influence of the dark-matter halo in the galaxy.  We briefly consider
signatures of possible megamagnifications throughout the history of the
solar system and in present-day neutrino observatories, and we remark
on lensing by nearby stars.

\section{Lensing Primer}
General Relativity predicts that a light ray (or neutrino) that
passes by a spherical body of mass $M$ is deflected by an
angle~\cite{Albert,PeacockCP,1987dmu..conf..133B}
\begin{equation}
\alpha=\frac{4GM}{c^2\xi}=\frac{2\mathcal{R}}{\xi}
\label{deflection}
\end{equation}
for an impact parameter $\xi$ much larger than the Schwarzschild radius
$\mathcal{R}= 2GM/c^2$, where $G = 6.6742 \times
10^{-11}\m^{3}\kg^{-1}\s^{-2}$ is Newton's constant and $c =
299\,792\,458\m\s^{-1}$ is the speed of light.  [Values not otherwise
attributed are taken from the \textit{Review of Particle
Physics}~\cite{PDBook}.]  For the special case of a ray that skims our
Sun, with impact parameter $\xi$ equal to the solar radius $R_{\odot} =
6.961 \times 10^{5}\km$ and $\mathcal{R}_{\odot} = 2.953\km$, the angle
of gravitational deflection is $\alpha = 8.48\murad$.  As we have
already noted, such a subtle deviation is unobservably small for
neutrinos.  The black hole at the center of the Milky Way galaxy, with
mass $M_{\bullet} = (3.61 \pm 0.32) \times 10^{6} M_{\odot} \approx 4
\times 10^{63}\gev$~\cite{Eisenhauer:2005cv} and Schwarzschild radius
$\mathcal{R}_{\bullet} = 1.07 \times 10^{7}\km \approx 3.46 \times
10^{-7}\pc$, is a better candidate for a neutrino lens.

\begin{figure}
\begin{center}
\includegraphics[width=8cm]{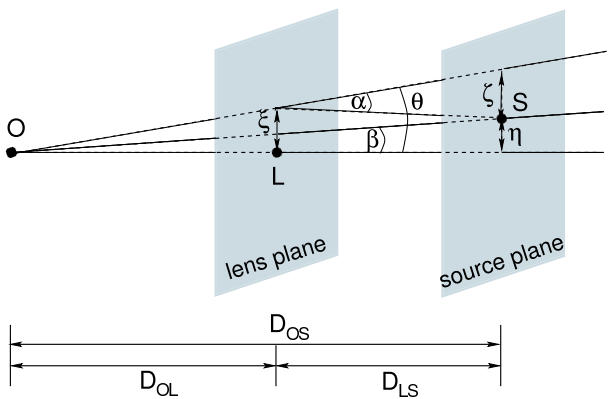}
\caption{Lensing geometry\label{fig:geometry}}
\end{center}
\end{figure}

In the simplest lensing set-up, shown in Figure~\ref{fig:geometry},
a compact lens of mass $M$ lies close to the line of sight
to a source, at a distance $D_{\mathrm{OL}}$ from the observer O.  The angle
$\beta$ describes the position of the source with respect to the lens
direction.  The angle $\theta$ describes the position of the apparent
source image with respect to the same axis.

If all the angles are very small, it is appropriate to define
angular-diameter distances $\eta=\beta D_{\mathrm{OS}}$, $\xi=\theta 
D_{\mathrm{OL}}$, and 
$\zeta=\alpha D_{\mathrm{LS}}$.  In this approximation, we can infer from Figure
\ref{fig:geometry} the lens equation,
\begin{equation}
\beta=\theta- \frac{\theta_{\mathrm{E}}^2}{\theta}\, ,
\end{equation}
where the Einstein angle $\theta_{\mathrm{E}}$ is 
\begin{equation}
\theta_{\mathrm{E}}=\sqrt{\frac{2 \mathcal{R} D_{\mathrm{LS}}}{D_{\mathrm{OS}}
D_{\mathrm{OL}}}}=\sqrt{\frac{2 \mathcal{R}}{D_{\mathrm{OL}}}\frac{x}{1+x}}\, ,
\label{thetae}
\end{equation}
with $x\equiv D_{\mathrm{LS}}/D_{\mathrm{OL}}$.  [For the interesting
case of the black hole at the center of the galaxy, {$8.0\kpc$} distant
from our location~\cite{1993ARA&A..31..345R}, the Einstein angle would be
{$\theta_{\mathrm{E}} = 3.1\murad$} for a source $1\kpc$ beyond
the galactic center, {$\thetae = 6.6\murad$} for a source 
opposite our location, and {$\theta_{\mathrm{E}} = 7.6\murad$} for
a source at the far edge of the galaxy ($x = 2$).]
In the plane OLS, the extremal angles of deflection are given by
\begin{equation}
\theta_{\pm}=\frac{\beta}{2}\pm\thetae
\sqrt{1+{\beta^2}/{4\thetae^2}}\, .
\label{image}
\end{equation}

Sufficiently strong lensing produces multiple images of the source.  If
the source, lens, and observer lie on a line ($\beta=0$), the multiple
images describe a perfect circle, or Einstein ring, with opening angle
$2\thetae$.  More generally, finite-size lenses give rise to a variety
of image patterns, depending on the particularities of the mass
distribution within the lens.  Distinct multiple images, arcs, or
Einstein rings have been observed for many light
sources~\cite{Wambsganss:1998gg}.  The separation between images---no
more than a few arcseconds---cannot be resolved in neutrinos.

In many cases the lens is not strong enough to produce multiple images,
arcs, or rings, but does create a distorted image of the source.
Usually the precise sizes and shapes of the sources are not known, but
it is possible to characterize \textit{average} properties.  The ``weak
lensing'' method~\cite{Bartelmann:1999yn}---comparing statistics of
sources and images---has been used to weigh nearby clusters by using
distributions of faraway galaxies as sources.  Weak lensing is also a
powerful tool in cosmology: statistics of the large scale structure
observed in distant galaxies or the cosmic microwave background make it
possible to infer the total (luminous plus dark) mass between the
observer and the source.  Large surveys---not in prospect for
neutrinos---are essential to the statistical reliability of inferences.

In the ``microlensing'' case~\cite{Roulet:1996ur}, multiple images are
overlaid too closely to be resolved and image distortion is too subtle
to observe, but light reaching the observer along multiple trajectories
increases the apparent brightness of the source.  Without knowing the
intrinsic brightness of the source, it is generally not possible to
determine the amplification, or ``magnification,'' as it is usually
called.  One can, however, observe the time variation of the brightness
of a source (such as a nearby star), as a heavy object passes between
source and observer.  Microlensing in this form is the basis of
searches for massive compact halo objects
(MACHOs)~\cite{Alcock:2000ph}.

The magnification is given by the ratio of solid angles in the presence 
and absence of the lens. At the extreme angles, we have
\begin{equation}
\mu_\pm=\frac{d\Omega\;\,}{d\Omega_0}=\abs{\frac{\theta_\pm \;d
\theta_\pm}
{\beta \;\;d\beta\;}}\, ,
\end{equation}
whereupon (cf.~(\ref{image}))
\begin{equation}
\mu=\mu_++\mu_-=
\frac{1+\cfrac{1}{2}\bbar^2}{\bbar\sqrt{1+ \cfrac{1}{4}\bbar^2}}\, 
\label{magnification}
\end{equation}
where $\bbar \equiv \beta/\thetae$ is the reduced misalignment angle.
The  magnification is maximized in the limit of perfect alignment,  as 
$(\beta, \bbar) \to 0$, for which
\begin{equation}
\mu \to \mu^{\mathrm{max}} = 1/\bbar\, .
\label{eq:mag}
\end{equation}
For a finite source of radius $R_{\star}$, the effective limit is
$\beta \to \beta_{\star} = {R_{\star}}/{D_{\mathrm{OS}}}$.  In that
limit, the magnification can be prodigious: for a source of radius
$R_{\star} = 10\km$ on the other side of the galaxy lensed by the black 
hole at the galactic center, we find 
\begin{equation}
\mu^{\mathrm{max}}=2.3 \times  10^{11} \sqrt{x(1+x)} \;.
\label{alignedmu}
\end{equation}
In comparison to the $\beta$ dependence of the amplification, the
dependence on the source location is of secondary importance: as $x$
varies between $0.01$ and $2$, $\mu^{\mathrm{max}}$ varies only by a
factor of $25$.

The calculation we have just made applies to a fictitious galaxy that
is empty except for the source, the black hole at the galactic center,
and the observer.  In the real Milky Way, matter obscures visible light
from the other side of the galaxy, but the magnification of gamma-ray,
radio, or neutrino sources might be observable. [We shall verify in 
\S\ref{sec:other} that diffuse 
matter throughout the galaxy contributes negligibly to lensing.]

Superposed lensed images arise from neutrinos that traverse different
paths, and so signals from the source reach the observer at different
times, as Krauss and Small have remarked~\cite{KraussSmall} for the
microlensing of light.  [The time delay between the arrival of neutrinos
traveling on different trajectories has been invoked in \cite{Barrow}
to explain a putative  bimodal time distribution of SN1987A neutrinos.]

It is convenient to calculate the transit time
along each geodesic by adding the interval from source to lens to the
interval from lens to observer.  The propagation time has two
components~\cite{Weinberg:GandC}: one corresponds to the time
required for straight-line propagation to and from the lens, and the
second is a general-relativistic time delay~\cite{PhysRevLett.13.789}
proportional to the Schwarzschild radius of the lens.  {For our
case of a black-hole lens, integration along the curved geodesic yields
the exact result,}
\begin{eqnarray}
ct&=& \sqrt{r_{\mathrm{O}}^2-\xi^2}+\sqrt{r_{\mathrm{S}}^2-\xi^2} \nonumber \\
& + &{\frac{\mathcal{R}}{2}}\left[
\sqrt{\frac{r_{\mathrm{O}}-\xi}{r_{\mathrm{O}}+\xi}}+2 \ln\left(\frac{r_{\mathrm{O}}+\sqrt{r_{\mathrm{O}}^2-\xi^2}}{\xi}\right) 
\right.\nonumber \\ [2mm]
& & \quad + \left. \sqrt{\frac{r_{\mathrm{S}}-\xi}{r_{\mathrm{S}}+\xi}}
+ 2 \ln \left(\frac{r_{\mathrm{S}}+\sqrt{r_{\mathrm{S}}^2-\xi^2}}{\xi}\right)
\right]
\end{eqnarray}
where $r_{\mathrm{O}}=D_{\mathrm{OL}}$ is the distance from the lens to
the observer, $r_{\mathrm{S}} = \sqrt{D_{\mathrm{LS}}^{2} +
\eta^{2}} = D_{\mathrm{OL}}\sqrt{x^{2} + \beta^{2}(1+x)^{2}}$ is the distance from the lens to the source, $\xi$ is 
the distance of closest approach to the lens, and 
$\mathcal{R}$ is the Schwarzschild radius of the lens. For a source 
$8\kpc$ beyond the galactic center, and with misalignment angle 
$\beta = \thetae$, the relativistic time delay induced by the black 
hole is $O(10^{3})\s$.

The longest geodesic corresponds to
$\xi_{-}=D_{\mathrm{OL}}|\theta_{-}|$ and the shortest one to
$\xi_{+}=D_{\mathrm{OL}}\theta_{+}$, with $\theta_\pm$ given in
Equation (\ref{image}).  Light signals emitted at the same time can
travel by different paths, so they reach an observer separated in
coordinate time by as much as
\begin{eqnarray}
\Delta t & \equiv & (t_{-} - t_{+}) \\
 & = & \frac{\mathcal{R}}{c}\left[2\bbar\sqrt{1 + \cfrac{1}{4}\bbar^{2}} +\ln{\left|\frac{1 + 
 \cfrac{1}{2}\bbar^{2} + \bbar\sqrt{1 + \cfrac{1}{4}\bbar^{2}}}{1 + 
 \cfrac{1}{2}\bbar^{2} - \bbar\sqrt{1 + \cfrac{1}{4}\bbar^{2}}}\right|}
 \right]\!. \nonumber
\end{eqnarray}
[An observer registers a proper interval $\Delta\tau =
\sqrt{g_{00}}\Delta t$; for an earthbound observer, $g_{00}$ differs
negligibly from unity.]  In the limit of very small misalignments
($\bbar \to 0$), the time dispersion is proportional to $\bbar$, viz.
\begin{equation}
    \Delta t \to \frac{\mathcal{R}}{c}\left[2\bbar +\ln{\left|\frac{1 + \bbar}{1 - 
    \bbar}\right|}\right] \approx \frac{4\mathcal{R}\bbar}{c} = 
    \frac{2\beta}{c} 
    \sqrt{2\mathcal{R}D_{\mathrm{OL}} \frac{1+x}{x}} \;,
\end{equation}
so that the magnification and time dispersion are reciprocally related through
\begin{equation}
    \mu\Delta t = 4\mathcal{R}/c \approx 142\s \; .
    \label{eq:reciprocity}
\end{equation}
In the small-$\bbar$ limit, both the magnification and the time spread
scale with the square root of the lens mass.  For the configuration
considered below Eqn.~{(\ref{eq:mag}), the time spread would be
{$\Delta t \approx 4.4 \times 10^{-10}\s$.} At the other extreme,
for $\bbar = 1$, {$\Delta t \approx 148\s$} and {$\mu =
3/\sqrt{5} = 1.34$.}

Figure \ref{fig:3d} shows the magnification and time dispersion as a
function of the misalignment angle for $\beta_{\star} \le \beta 
\lesssim \thetae$.
\begin{figure}
\begin{center}
\includegraphics[width=8cm]{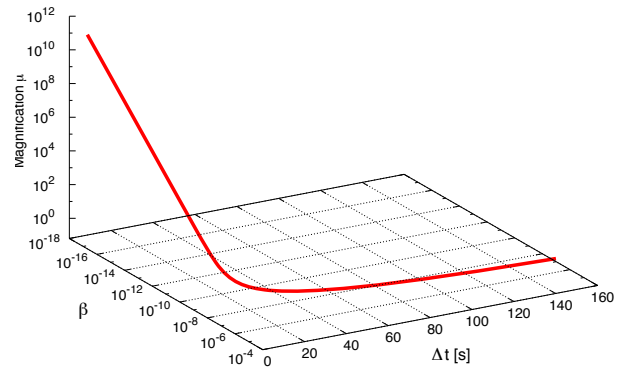}
\caption{Magnification $\mu$ and dispersion in time $\Delta t$ as a
function of misalignment angle $\beta$, for the case of a 10-km-radius
source opposite our location in the galaxy.  \label{fig:3d}}
\end{center}
\end{figure}
For $\beta \ll \thetae$, we observe the $\mu \propto \beta^{-1}$ 
behavior of Eqn.~(\ref{eq:mag}). As $\beta$ increases toward $\thetae$, the 
magnification diminishes and the dispersion in time increases to many 
seconds.

\section{Galactic Supernova Lensed by the Black Hole at the Galactic Center 
\label{sec:GalSN}}
\subsection{General Characteristics \label{subsec:genchar}}
The lensing phenomena we have described for light signals (in a
hypothetical nearly empty galaxy) apply essentially unchanged for
neutrinos, with utterly negligible corrections for the neutrino
velocity, which is slower than the speed of light by
$\cfrac{1}{2}(m_{\nu}/p_{\nu})^{2}\cdot c$.  The Milky Way galaxy is
highly transparent to low-energy neutrinos.  The $\nu N$ interaction
length at $E_{\nu} = 10\mev$, for example, exceeds $10^{17}\cmwe$,
{whereas an average diameter through the galactic disc integrates
a column density considerably less than
$1\cmwe$~\cite{Berezinsky:1992wr}.} Subtle effects due to neutrino
lensing by stars (including the Sun) and galactic halos have been
identified in Ref.~\cite{Escribano:2001ew}.  Here we analyze a more
dramatic illustration of gravitational lensing: A supernova at superior
conjunction to the black hole at the galactic center would be an ideal
source for the study of gravitational lensing of neutrinos.

The gravitational binding energy $E_{B} =3 \times 10^{53}\hbox{ erg}$
of a core-collapse supernova is roughly equipartitioned among the six
neutrino and antineutrino species.  A useful description of the
supernova luminosity in neutrinos consists of a nearly instantaneous
rise followed by an exponential decay~\cite{Fogli:2003dw,Beacom:1998fj}
that can be represented by
\begin{equation}
L_0(t) = \frac{E_{B}}{6 \tau}\exp^{-t/\tau} \equiv L_{0}\exp^{-t/\tau}.
\label{standardl}
\end{equation}
The decay time $\tau = 3\s$ implies an effective pulse length
of $\approx 10\s$, consistent with SN 1987A observations. 

A time spread of neutrinos arriving at Earth from a lensed 
supernova that considerably exceeds the canonical 10-s pulse length 
would be a signature of lensing. We plot in Figure~\ref{fig:timespectrum} 
some examples of the time profiles to be expected for a supernova at 
distance $D_{\mathrm{OS}} = 16\kpc$, for some representative values of 
the misalignment angle $\beta$.
\begin{figure}[t]
\begin{center}
\includegraphics[width=8.5truecm]{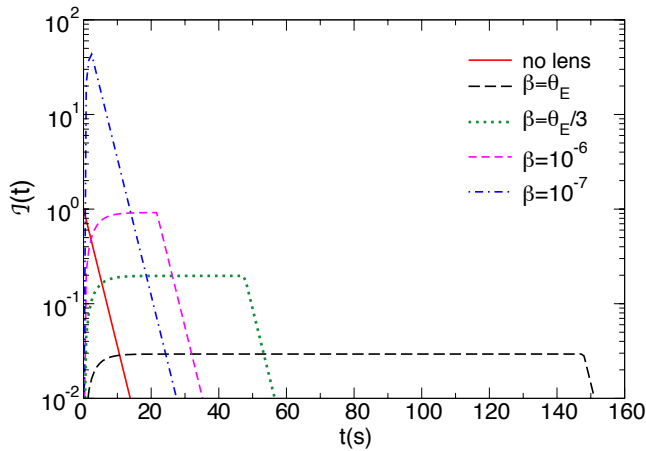}
\caption{Normalized neutrino intensity as a function of time 
for lensed supernovae with $\beta = \thetae$ (black dashed curve),
$\beta = \thetae/3$ (green dotted),
$\beta=10^{-6}$ (magenta short dashes), $\beta=
10^{-7}$ (blue dot-dashed). The no-lensing case is shown as the 
red solid curve.}
\label{fig:timespectrum}
\end{center}
\end{figure}

In the absence of lensing, the intensity of neutrinos arriving at Earth
as a function of time is an exponentially decaying pulse proportional
to $L_{0}(t)$.  When the supernova neutrinos are lensed, neutrinos may
arrive by different paths, with different travel times.  For very small
misalignment angles, the difference in path lengths is very small, and
so the resulting intensity profile is indistinguishable from $\mu
\times$ the unlensed profile.  An example is given in the curve
corresponding to $\beta = 10^{-7}$ in Figure~\ref{fig:timespectrum},
which is very nearly given by $\mu = 65.7$ times the unlensed intensity
profile.  The normalization of the intensity curves $\mathcal{I}(t)$
plotted in Figure~\ref{fig:timespectrum} is such that
$\int_{0}^{\infty} dt \,\mathcal{I}(t) = \mu\tau$.

As the misalignment angle $\beta$ increases toward $\thetae$, new and
longer paths contribute, so the intensity is nearly constant over an
increasingly long time interval.  Thereafter, no new paths come into
play and the intensity decays according to the exponential in
$L_{0}(t)$, but displaced in time.  The onset of this distortion of the
pulse is shown by the $\beta = 10^{-6}$ curve in
Figure~\ref{fig:timespectrum}; the fully developed behavior is
exhibited for the case $\beta = \thetae$.  For values of $\beta \approx
\thetae/3$, both the magnification and the time delay are significant.

\subsection{Likelihood of lensing events \label{subsec:likelihood}}
The near-perfect alignment of source, lens, and observer is a very 
special circumstance. How frequently might a dramatic lensing event 
occur?

Let us first assume that supernovae are distributed according to the 
mass density in the galactic disc~\cite{Gates:1995js},
\begin{equation}
    \sigma(r) = \sigma_{0} e^{-r/r_{0}}\hbox{, for }r \le r_{G}\;,
    \label{eq:lumden}
\end{equation}
where $r$ is the distance from the galactic center, $r_{G} = 15\kpc$ is
the galactic radius, and the parameter $r_{0} = 3.5\kpc$.  [It is an
excellent approximation for our purposes to idealize the disc as
infinitely thin, so that $\sigma(r)$ measures the luminous mass density
per unit area.]  A useful measure of the fraction of supernovae that
lie along a swath at radius $r = D_{\mathrm{LS}} = x D_{\mathrm{OL}}$
is then
\begin{equation}
    f(r) = \frac{2\pi\sigma(r) r \Delta r}{2\pi \int_{0}^{r_{G}} dr 
    r \sigma(r)}\;,
    \label{eq:rswath}
\end{equation}
where $\Delta r = 2R_{\star} \approx 20\km$ is a typical supernova 
diameter. In terms of the dimensionless quantities $x$ and $x_{0} = 
r_{0}/D_{\mathrm{OL}} = 3.5/8$, we find
\begin{equation}
    f(x) = 4.6 \times 10^{-16} \,x e^{-x/x_{0}}\;.
    \label{eq:xswath}
\end{equation}

Only a tiny fraction of the supernovae that occur at a given radius 
are aligned closely enough with observer and lens---with a detector on 
Earth and the black hole at the center of the galaxy---to be 
observably lensed. A reasonable measure is the ratio 
$\thetae/\pi$,
which leads to the probability that a galactic supernova at reduced 
radius $x$ be lensed toward Earth of
\begin{equation}
    P(x) = f(x) \cdot \frac{\thetae}{\pi} \approx 1.35 \times 
    10^{-21}\,x e^{-x/x_{0}}\sqrt{\frac{x}{1+x}}\;.
    \label{eq:swathprob}
\end{equation}
Integrating over all radii, we find that the fraction of supernovae 
lensed toward Earth is approximately $1.8 \times 10^{-6}$. 

It has been argued that while the distribution of Type IA supernovae
 closely tracks the luminous matter, the distribution of neutron stars 
 or pulsars is a better tracker of the core collapse
supernovae that concern us here~\cite{Ferriere:2001rg}. Under either 
assumption, the density has the form
\begin{equation}
    \sigma(r) = \sigma_{0} r^{p}\,e^{-r/r_{0}}\hbox{, for }r \le r_{G}\;,
    \label{eq:NSden}
\end{equation}
where ($p = 4$, $r_{0} = 1.25\kpc$) for neutron stars and ($p = 2.35$, 
$r_{0} = 1.528\kpc$) for pulsars. The radius at which the density of 
supernovae is greatest is approximately $5\kpc$ for the neutron-star 
model and $3.5\kpc$ for the pulsar model.

If core-collapse supernovae track the neutron-star distribution, we 
compute the
fraction of supernovae that lie along a swath at radius $r =
D_{\mathrm{LS}} = x D_{\mathrm{OL}}$ as before:
\begin{equation}
    f(x) =4.7\times 10^{-14} \, x^5 e^{-x/x_0}\;,
    \label{eq:fracneutstar}
\end{equation}
where now $x_{0} = r_{0}/D_{\mathrm{OL}} = 1.25/8$.
Now the probability that a galactic supernova at reduced 
radius $x$ be lensed toward Earth is
\begin{equation}
P(x) \approx 1.4\times 10^{-19} \,x^5
e^{-x/x_0}\sqrt{\frac{x}{1+x}}\;.
\label{eq:NSProb}
\end{equation}

The fraction of galactic supernovae lensed toward Earth is 
$2 \times 10^{-6}$, quite comparable to the fraction we found if the 
 supernovae are assumed to  follow the visible matter 
distribution. [The fraction for the pulsar distribution is $1.9 \times 
10^{-6}$.]
We show in Figure~\ref{fig:muprob} the probability 
that a lensed supernova be amplified by a factor $\mu$ or greater,
if the supernovae sites track the distribution of neutron stars. 
Our other two hypotheses yield very similar results.
\begin{figure}[t]
\includegraphics[width=9truecm]{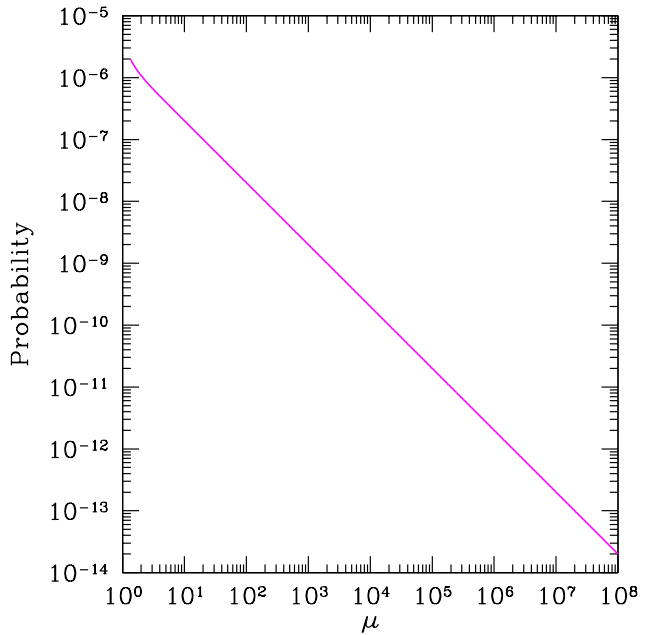}
\caption{Probability that lensing of a core-collapse supernova by the 
black hole at the center of the galaxy amplifies the neutrino flux at 
Earth by a factor $\mu$, assuming that the distribution of supernovae 
tracks the population of neutron stars in the galactic disc.}
\label{fig:muprob}
\end{figure}


The current rate of galactic supernovae that produce neutrino bursts is
estimated~\cite{Tammann:1994ev} at one in $47 \pm 12\y$.  If the rate
of supernova neutrino bursts throughout the galaxy is constant over
time at the current rate of approximately two per century, we conclude
that on the order of 180 lensed supernovae have occurred throughout the
4.5-Gy history of the solar system.  The time between lensing events
with amplification factor $\mu = 10$ is approximately 250 million
years, and an event with $\mu = 100$ has occurred perhaps once in the
history of the solar system.  The frequency of events with
magnifications $\mu \ge 10$ is comparable to the rate of nearby
supernova
explosions~\cite{Ruderman:1974,Ellis:1993kc,Ellis:1995qb,Fields:1998hd,Collar:1995mh}.
These are conservative estimates, given that the supernova rates are
not constant over time and  were apparently higher when the galaxy 
was young~\cite{Strigari:2005hu}.

It is also interesting to ask how many supernovae exhibit a noticeably
increased pulse-length as a consequence of gravitational lensing.  The
fraction of supernovae for which the time dispersion lies between
$50\s$, for which $\mu = 2.99$, and the value ($148\s$, with
$\mu = 1.34$) corresponding to $\beta = \thetae$ is $1.3 \times
10^{-6}$.  Over the history of the solar system, supernova neutrino
pulses longer than $50\s$ have arrived at Earth approximately 117
times---an average of one every 38.5 million years. Lowering the 
requirement to a time dispersion of $20\s$, for which $\mu = 7.16$, 
increases the fraction to $1.72 \times 10^{-6}$. This rate 
corresponds to 155 stretched supernova neutrino pulses over the age 
of the solar system, roughly one every 29 million years.

\subsection{Signatures in Neutrino Observatories}
A few dozen $\bar{\nu}_e$ from SN1987A were recorded in a number of
detectors (IMB~\cite{PhysRevLett.58.1494}, Kamiokande
II~\cite{PhysRevLett.58.1490}, Baksan~\cite{1987sn...work..237A}, and
Mont Blanc~\cite{1987EL......3.1315A}) via the charged-current process
$\bar{\nu}_e + p \rightarrow e^{+} + n$.  These observations set the
stage for the detection of neutrinos from future supernovae.  [It is
less certain~\cite{Costantini:2006xd} that $\nu_{e}$-initiated events
associated with SN1987A have been established.  The detection of a
single in-time $\nu_{e}e$ event would test the equivalence principle
for $\nu_{e}$ \textit{vs.} $\bar{\nu}_{e}$ to a few parts in a
million~\cite{PhysRevD.38.3313,PhysRevD.39.1761}.]

The signatures of a core-collapse supernova lensed by the black hole 
at the galactic center are a significant amplification of the 
neutrino flux at the detector and a dispersion in time of the neutrino 
burst. The larger each of these effects---which are correlated if 
lensing occurs---the smaller is the likelihood that the supernova is 
simply an outlier. Pointing information derived from neutrino 
interactions is crude, but to implicate lensing it would suffice to 
identify the supernova direction along a line through the galactic 
center.

The yield of neutrinos emitted by a core-collapse supernova can be 
anticipated within a factor of two to three~\cite{Raffelt:2002tu}. 
Under that assumption, a measurement of the distance to the supernova 
provides a good estimate of the unlensed neutrino flux. For supernovae 
opposite our location in the galaxy, it is likely that visible light 
would be so attenuated traversing the matter in the galactic disc that 
the precursor star could not be identified. The Chandra mission, with 
its angular resolution of $2.5\murad$, has catalogued many x-ray point 
sources toward the galactic center~\cite{Muno:2003ne}. 

By measuring the neutrino flux alone, one could infer an apparent distance to
an unlensed source.  With a flux greatly magnified by lensing, the
apparent distance might be implausibly small, placing the supernova on
this side of the galactic center, where the optical signal would have
been visible.  Lensing would then be strongly implicated, even without
a precise determination of the magnification.

We summarize in Table~\ref{tab:tabdetectors} some characteristics for
supernova neutrino detection of the available techniques, which entail
several elements: sensitivity to multiple neutrino flavors (through
neutral-current measurements), timing, energy resolution, and pointing
back to the source (through $\nu e$ elastic
scattering~\cite{Beacom:1998fj}).  The most interesting detector
capabilities for supernova lensing applications are the time resolution
and the capability to point to the source.  All of the techniques
listed in Table~\ref{tab:tabdetectors} allow for an adequate
measurement of arrival time.  We will focus on water Cherenkov
detectors (such as Super-Kamiokande~\cite{Fukuda:2002uc}) and
long-string ice/water Cherenkov detectors (such as
AMANDA~\cite{Andres:1999hm}), because they are operating and the
next-generation version of AMANDA (IceCube~\cite{Montaruli:2006eg}) is
under construction.

It is convenient to scale the total number of neutrino events in a 
 detector of given effective mass as~\cite{Beacom:1998ya}
\begin{equation}
N=N_{0}\left(\frac{E_B}{3 \times 10^{53}\hbox{ erg}} \right)
\left(\frac{10\kpc}{D_{\mathrm{OS}}}\right)^2~, \label{eq:nevents}
\end{equation}
where as usual $E_{B}$ is the gravitational binding energy of the
collapsing star and $D_{\mathrm{OS}}$ is the distance between observer
and source.  Assuming sensitivity to all reactions, the reference rate
is $N_{0} \approx O(10^{4})$ for the Super-Kamiokande detector with 32
kton of $\mathrm{H_{2}O}$ and a 5-MeV threshold.  [For the SNO
detector~\cite{Boger:1999bb}, which is expected to stop running at the
end of 2006, $N_{0}\approx O(10^3)$.]  Therefore,  a supernova with
a typical 10-s time profile located at the symmetrical configuration
($D_{\mathrm{OS}}= 16\kpc$)  would generate approximately $3\,900$
total events in a Super-Kamiokande--like detector---in the absence of
lensing effects.

If supernova neutrinos are lensed by the black hole at the center of
our galaxy, the number of events in a Super-Kamiokande class detector
could be ten times greater, with a time dispersion of $15\s$, or five
times greater, with a time dispersion of $30\s$, depending on the value
of the misalignment angle $\beta$, see {Figure~\ref{fig:3d}.} The
neutrino time profile would be greatly distorted from what would be
reconstructed in the absence of lensing: it would be delayed, highly
amplified, and broadened, as can be seen from Figure
\ref{fig:timespectrum}.  It is worth noting that Super-Kamiokande can
measure up to $30\,000$ events within the first second of a burst, with
no dead time~\cite{Fukuda:2002uc}.

Among several proposals for future water Cherenkov detectors, UNO (the
Ultra underground Nucleon decay and neutrino
Observatory)~\cite{Jung:1999jq}, with a fiducial volume twenty times
Super-Kamiokande's, would record $10^5$ neutrino interactions from a
supernova located $10\kpc$ away, which means $39\,000$ events if
$D_{\mathrm{OS}} = 16$ kpc.  In the presence of lensing effects, the
number of events recorded in UNO could reach an order of magnitude 
larger, $4 \times 10^5$ events with a time dispersion of $15\s$, 
or  five times larger, $2 \times 10^5$ events dispersed over $30\s$.  

Although the energy of supernova neutrinos lies far below the threshold
for track reconstruction in long-string Cherenkov detectors, a supernova
neutrino burst would give rise to a coincident rate increase above
ambient noise in all photomultipliers in a large array.  The 
shortcoming of this technique is that it gives no information about 
the direction to the supernova.

IceCube, the successor to AMANDA, is under construction at the South
Pole.  A supernova $10\kpc$ away would generate $1.5\times 10^{6}$
excess photoelectrons over $10\s$ in IceCube's $4\,800$ optical
modules, to be compared with a background noise of $1.44 \times 10^{7}$
photoelectrons, for a favorable signal-to-noise ratio $S/\sqrt{N} \approx
400$~\cite{Dighe:2003be}.  If we rescale to our ideal situation
$D_{\mathrm{OS}}=16\kpc$, the number of excess photoelectons would be
$6\times10^5$, for which $S/\sqrt{N} \approx 160$.  In the presence of
lensing, both $S$ and $S/\sqrt{N}$ could be considerably enhanced.  The
standard supernova searches in AMANDA/IceCube bin data on 500~ms-10~s
scales because of the typically low signal-to-noise ratio.  For a
lensed supernova with enhanced signal-to-noise, the AMANDA supernova
detection system, which records data over intervals as short as 10~ms,
could reconstruct a very precise time spectrum.  
  
\begin{table*}[t]
    \caption{Neutrino observatory techniques and their capabilities 
    for the detection of supernova neutrinos.}
    \label{tab:tabdetectors}
\begin{ruledtabular}
\begin{tabular}{cccc}
 Detector type: Examples & Energy resolution  & Pointing & Flavor 
 tag \\
\hline
Water Cherenkov: Super-Kamiokande~\cite{Fukuda:2002uc}, UNO~\cite{Jung:1999jq}
& Yes  & Yes & $\bar{\nu}_e$\\
\hline  
$\begin{array}{c} \hbox{Long-string Cherenkov: AMANDA~\cite{Andres:1999hm}, 
IceCube~\cite{Montaruli:2006eg},} \\ \hbox{Baikal~\cite{Lubsandorzhev:2005fe}, NESTOR~\cite{Resvanis:2006sb}, 
ANTARES~\cite{Flaminio:2005rw}} \end{array}$ 
& No  & No & $\bar{\nu}_e$\\
\hline
Scintillator: KamLAND~\cite{Suzuki:1999dg}, LVD~\cite{Bari:1989nw} & Yes & 
No & $\bar{\nu}_e$\\
\hline  
Heavy water: SNO~\cite{Boger:1999bb} & $\begin{array}{c} \hbox{Yes} \\ 
\hbox{No} 
\end{array}$  & 
$\begin{array}{c} \hbox{Yes} \\ \hbox{No} \end{array}$ & 
$\begin{array}{c} \bar{\nu}_e\hbox{ (CC)} \\ \hbox{all (NC)} \end{array}$\\
\hline  
High-$Z$ / neutron: OMNIS~\cite{Zach:2002is}, LAND~\cite{Hargrove:1996zv} & 
Yes  & No & all\\
\hline  
Liquid Argon: ICANOE~\cite{Rubbia:1999py} & Yes  & Yes & 
{$\nu_e, \bar{\nu}_{e}$}\\
\end{tabular}
\end{ruledtabular}
\end{table*} 

\subsection{Signatures in the Historical Record}
Because the probability of witnessing a lensed supernova is so tiny, it
is worthwhile to ask whether it might be possible to recognize the
effects of a past event, and so greatly increase the integration time
for observations.  Two possibilities come to mind: isotopic anomalies
and doomsday events.  Neither seems promising as an unambiguous marker
for a supernova lensed toward Earth.

Supernova neutrinos may induce inverse beta decay in nuclei within the
Earth.  Indeed, the idea of radiogeochemical
studies~\cite{Haxton:1987bf} such as the molybdenum-technetium
experiment~\cite{Cowan:1981wg,Wolfsberg:1984sd} is to
infer the supernova rate in the galaxy by integrating the neutrino flux
over several millions of years.  It exploits a reaction on a naturally
occurring ore target that leads to an isotope with a half-life far
shorter than the age of Earth,
\begin{equation}
    \nu + \nucl{Mo}{98}   \to   \nucl{Tc}{97}~(t_{1/2} = 2.6\my)  + n + 
    e^{-} \; ,
\label{eq:molyb}
\end{equation}
for which the 7.28-MeV threshold excludes most solar neutrinos.

Haxton \& Johnson~\cite{Haxton:1987bf} have found that the average flux of neutrinos from 
galactic supernovae is reproduced by placing all supernovae at 
$4.6\kpc$ from Earth. The neutrino flux from a lensed supernova at 
distance $8\kpc \cdot (1+x)$ is thus
\begin{equation}
    \mathsf{R} = \frac{\mu}{(1+x)^{2}} \left( \frac{4.6\kpc}{8\kpc} 
    \right)^{2}
    \label{eq:lensoverdiffuse}
\end{equation}
times that from a prototypical supernova. The Mo-Tc technique integrates neutrino
fluxes over several million years.  Over $1\my$, at the current rate 
of 2 galactic supernovae per century, the ratio of neutrinos from a 
single lensed supernova to the unlensed galactic background is 
\begin{equation}
    \bar{\mathsf{R}} = \frac{\mathsf{R}}{2 \times 10^{4}} = 1.65 
    \times 10^{-5} \frac{\mu}{(1+x)^{2}}.
    \label{eq:megarat}
\end{equation}
Now, for a single, perfectly aligned event, we would find using 
Eqn.~(\ref{alignedmu}) a megayear 
ratio
\begin{equation}
    \bar{\mathsf{R}} = 3.8 \times 10^{6} \sqrt{\frac{x}{(1+x)^{3}}} 
    \approx 1.34 \times 10^{6}, \hbox{ for }x=1.
    \label{eq:alignedmegarat}
\end{equation}
A supernova opposite our location in the galaxy would stand out 
dramatically from the unlensed background.

On the other hand, we have seen that highly magnified supernovae are
exceedingly rare, with only a single $\mu \gtrsim 100$ event having
occurred over the lifetime of the solar system.  A magnification $\mu
\gtrsim 2 \times 10^{6}$ would cause a 3-$\sigma$ excess in the megayear
integrated neutrino flux.  Such an event is stupendously improbable,
occurring only for one galactic supernova in $10^{12}$.

Even under such propitious circumstances, the factor-of-three
uncertainty in the rate of galactic supernovae, the existence of
backgrounds from $\nu_{\mathrm{solar}} \nucl{Mo}{97} \to \nucl{Tc}{97}
+ e^{-}$, and the possibility that nearby star-forming regions might
constitute an uncontrolled background~\cite{Nguyen:2005dq}, would make
it challenging to attribute a larger-than-nominal \tnucl{Tc}{97}
abundance to a lensing event.  A possible discriminant might be found
in other isotopic abundances~\cite{Ellis:1995qb,Fields:1998hd} that
might carry the imprint of cosmic rays associated with a nearby
supernova explosion.

It is natural to wonder whether a nearby supernova might have triggered
an extinction episode in the paleontological record.  Ellis \&
Schramm~\cite{Ellis:1993kc} concluded that the $\gamma$ fluxes and
charged cosmic rays from a supernova explosion at $10\pc$ from
Earth---a once in a few hundred million years occurrence---would
produce a hole in the ozone layer that would admit lethal solar
radiation, dramatically altering photosynthesis cycles.  The ionizing 
radiation that would cause such a cataclysm would not reach the solar 
system in significant amounts if the supernova were located instead at 
the other side of the galaxy.

Whether supernova neutrinos would provoke appreciable damage to living 
organisms is a matter of debate. At issue is whether nuclei recoiling 
from neutrino collisions would cause irreparable damage to genetic 
material~\cite{Collar:1995mh}. Our only comment is that if neutrinos 
from a nearby  supernova did have lethal effects, the neutrino flux 
from a perfectly aligned supernova opposite our location would be more 
damaging still. The huge magnification factor of $\mu \approx 3 \times 
10^{11}$ more than compensates the $r^{-2}$ supression,
making the neutrino flux up to five orders of magnitude greater than
in the case of a supernova $10\pc$ from Earth.

\section{Other Lenses and Sources \label{sec:other}}
Under the rare circumstances of a supernova opposite our location in
the galaxy, lensing of neutrinos by the black hole at the galactic
center could produce a prodigious magnification of the neutrino flux at
Earth.  It is worth considering whether other microlensing
effects---induced, for example, by multiple gravitational scattering on
the dark matter in the galaxy---might diffuse the supernova neutrinos
and so diminish the magnification.  These effects are negligible for
the case at hand.

The dark matter that could disrupt our conclusions about black-hole
lensing is essentially contained within the extremal neutrino geodesics
that we consider.  To obtain a slightly generous estimate, we compute
the dark matter contained within cones of opening angle $\thetae$,
directed from source and observer toward the galactic center.  For the
halo profiles given in
Refs.~\cite{Ascasibar:2005rw,Binney:2001wu,Navarro:1996gj,Moore:1999nt},
dark matter contained in the ``geodesic cones'' is less than one
percent of the black-hole mass, an amount that may safely be neglected.

Stars are effective lenses for light, which is observed with excellent
angular resolution, but are unlikely to produce observable effects for
neutrinos~\cite{Escribano:1999gy,Escribano:2001ew}, even when
finite-lens effects are taken into account.

For the lensing of a distant source by a nearby star, 
$D_{\mathrm{OL}}$ is relatively small, while $x=D_{\mathrm{LS}}/D_{\mathrm{OL}}$ is
very large.  The magnification for small misalignment angle $\beta$ is 
\begin{equation}
\mu\approx\frac{\theta_E}{\beta}\simeq\frac{\sqrt{2 \mathcal{R}
D_{\mathrm{OL}}}}{R_{\star}}x\, , 
\label{eq:starmag}
\end{equation}
where $\mathcal{R}$ is the Schwarzschild radius of the star and
$R_{\star}$ is the radius of the source.  For the case of a
core-collapse supernova lensed by $\alpha$Cen, approximately one solar
mass $1.34\pc$ distant from Earth, $\mu \approx 10^{6}x$.  For a
supernova at a distance of $1\mpc$ the magnification would be again
enormous, $\mu \approx 10^{12}$.  An unlensed supernova at this
distance for which one detected neutrino event would be
expected~\cite{Ando:2005ka} would result in some $10^{12}$ events if
perfectly aligned with $\alpha$Cen and Earth.  It is thus conceivable
that lensing could allow for the detection of significant numbers of
neutrinos from more distant supernovae.  However, the Einstein angle
for $\alpha$Cen is nearly twenty times smaller than that for the
black-hole scenario, so the probability of perfect alignment is
proportionately reduced.  No extreme magnification events have
been observed for light.

Many other potential neutrino sources, including extremely distant
supernovae, gamma ray bursters, and active galactic nuclei, could in
principle be greatly magnified because $x$ is so large for a distant
source and nearby lens.
The number of such neutrinos reaching Earth from extragalactic sources scales as
\begin{equation}
 \frac{\mu}{D_{\mathrm{OS}}^2}=\frac{\mu}{D_{\mathrm{OL}}^2 (1+x)^2} 
 \propto \frac{\mu}{x^{2}}\, ,
\end{equation}
which implies low counting rates in the absence of lensing for sources
at many megaparsecs.  As can be seen from Eq.  (\ref{eq:starmag}), the
magnification compensates one power of $x$. 

Only for supermassive lenses might the magnification be large
enough to overcome the small anticipated flux from a very distant
source.  We know of no nearby candidates as effective as the black hole
at the galactic center. Magnification by distant lenses (for which
$D_{\mathrm{OL}}$ is very large) is likely to be modest, because $x$ is
small in that case.

To summarize, the distribution of dark matter within our galaxy is too
sparse to act as a diffusing lens that significantly diminishes the
amplification of neutrino flux for a supernova lensed by the black hole
at the galactic center.  Moreover, the setup of a galactic supernova
lensed by the galactic center black hole seems the most promising
configuration for the observation of gravitational lensing of
neutrinos.  In most other configurations for which the neutrino flux
might be highly amplified, lensing would also be observed in light,
whereas supernovae on the other side of the galactic center are only
visible in neutrinos.

\section{Conclusions \label{sec:conc}}
Observation of lensed neutrinos emitted by a core-collapse supernova 
would constitute a graphic demonstration of the gravity's influence on 
neutrinos. Amplification of the neutrino flux at Earth by many orders 
of magnitude may occur for near-perfect alignment of supernova, black 
hole, and Earth, but such events are exceedingly rare. We estimate 
that a lensing event with magnification by two orders of magnitude has 
occurred once in the history of the solar system, and the mean time 
between factor-of-ten events is 250 million years. 

A dispersion of the arrival time of supernova neutrinos is a slightly more
promising marker for not-quite-perfect alignment.  A pulse stretched by
more than $20\s$ has occurred on average once in 29 million years.
Neutrino telescope observers should be alert to the interesting
possibility of distorted profiles in time for neutrino bursts that
emanate from beyond the galactic center.

For all of this, the observation of a spectacular lensing of supernova
neutrinos in real time is highly improbable.  While ``it is a part of
probability that many improbable things must happen~\cite{Agathon},''
we shall have to look elsewhere to make quantitative studies of the
gravitational interactions of neutrinos.

\acknowledgments
We thank John Beacom, Tyce DeYoung, Evalyn Gates, Nick Gnedin, and Joe
Lykken for helpful discussions.  This work was supported in part by
National Science Foundation grant PHY-0555368.  Fermilab is operated by
Universities Research Association Inc.\ under Contract No.\
DE-AC02-76CH03000 with the U.S.\ Department of Energy.  This work was
supported in part by the European Programme ``The Quest for
Unification.''  contract MRTN-CT-2004-503369.

\bibliography{nulensrefs}

\end{document}